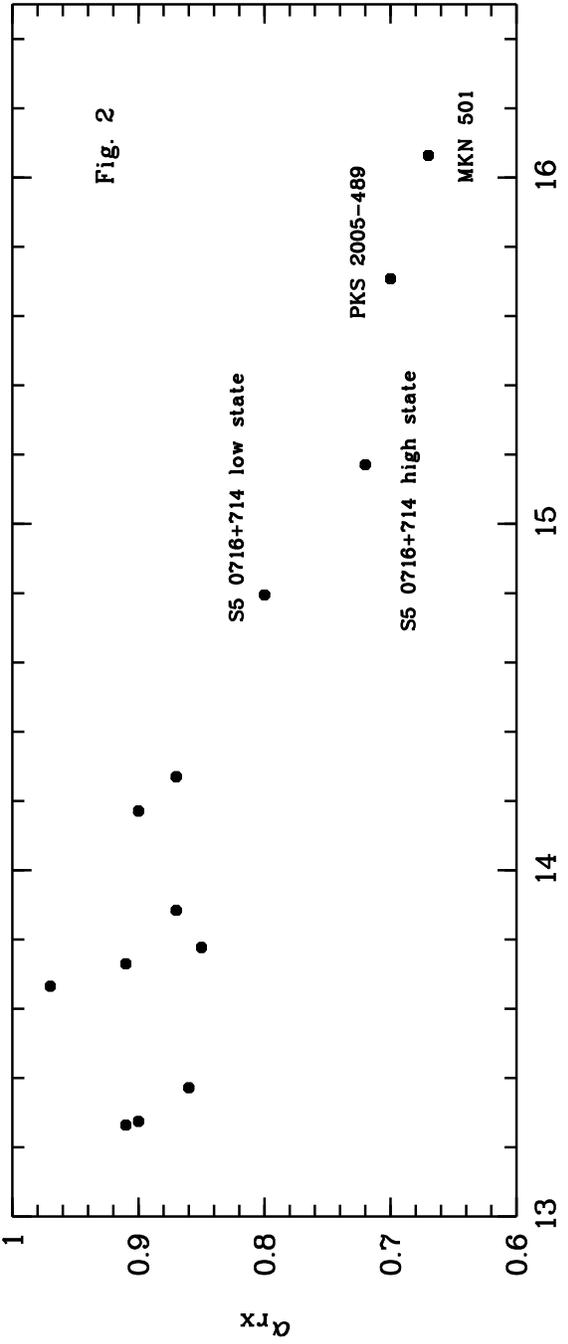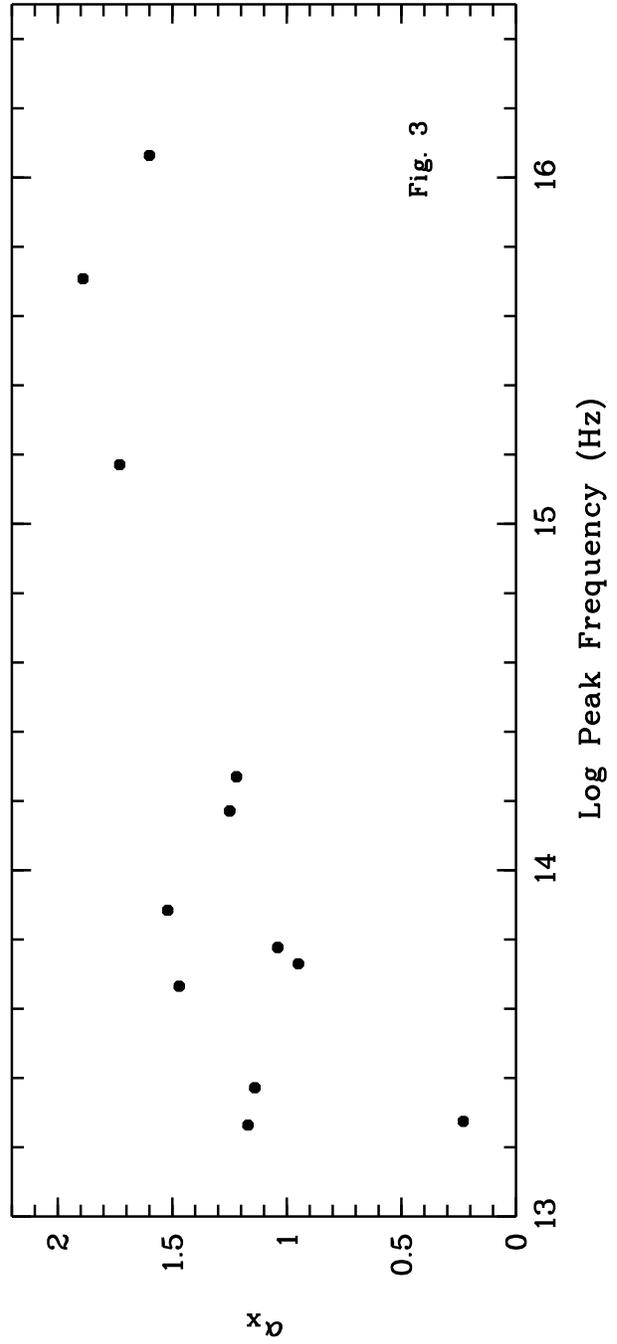

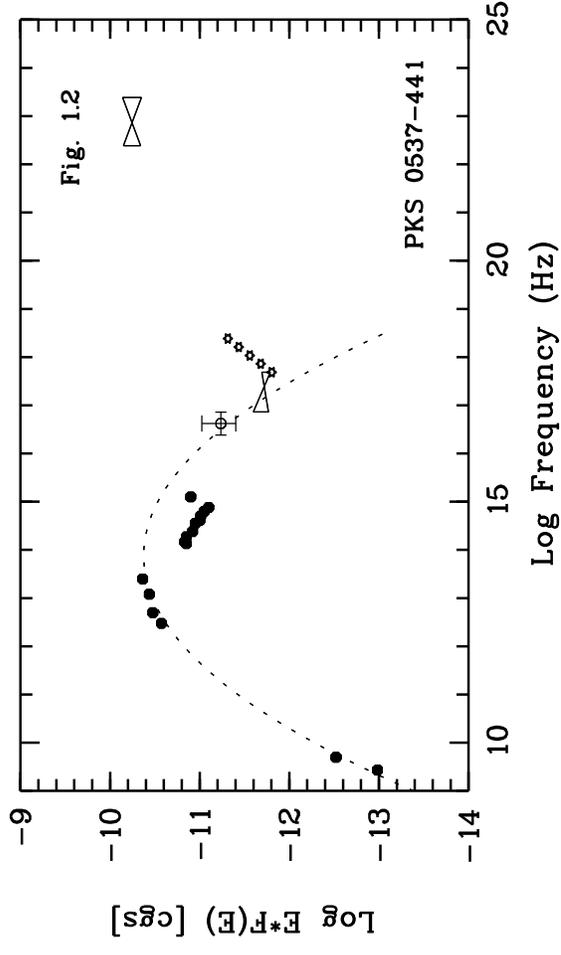
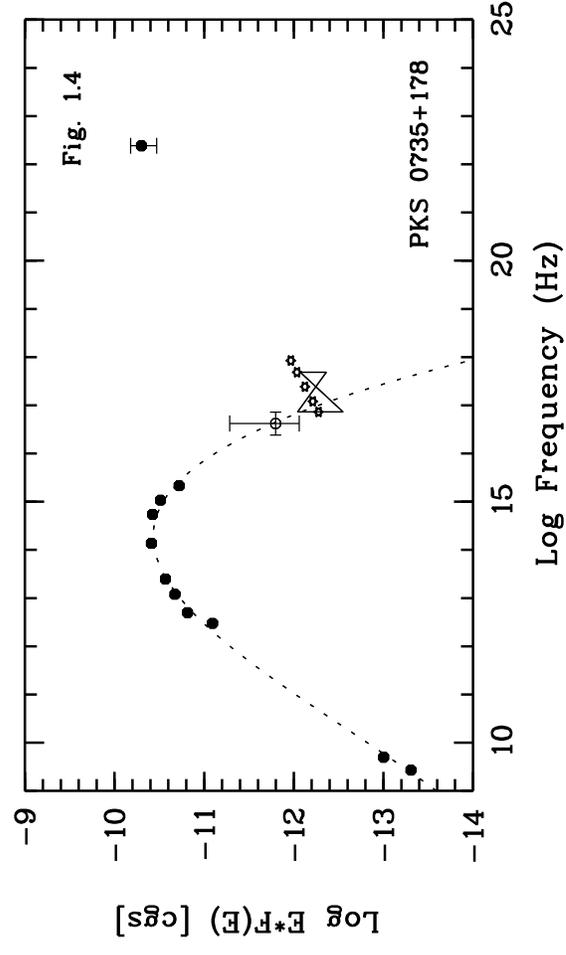
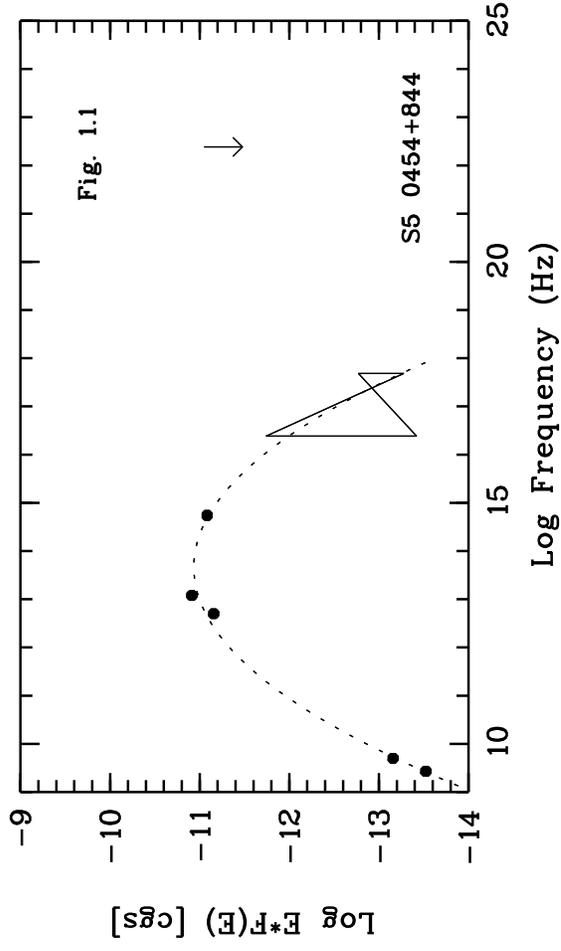
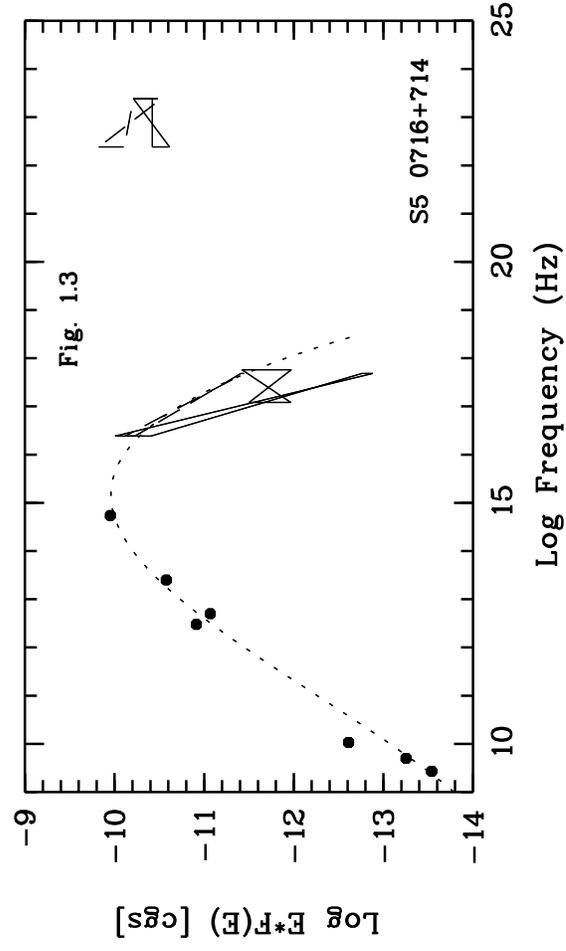

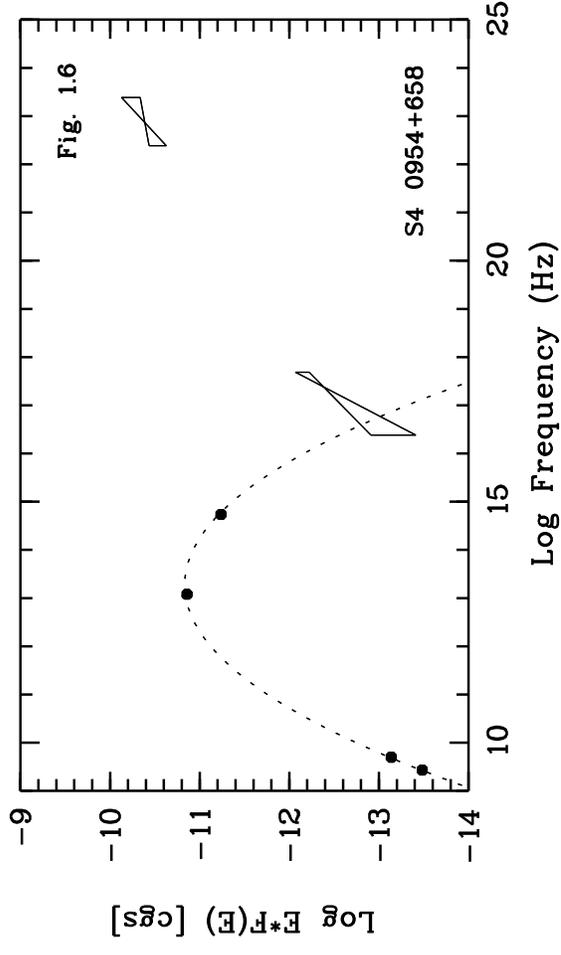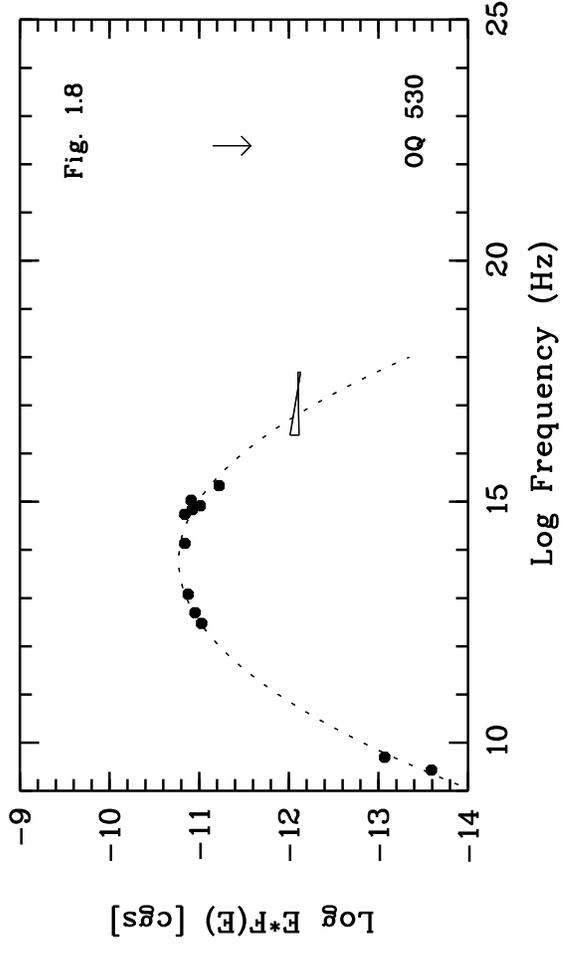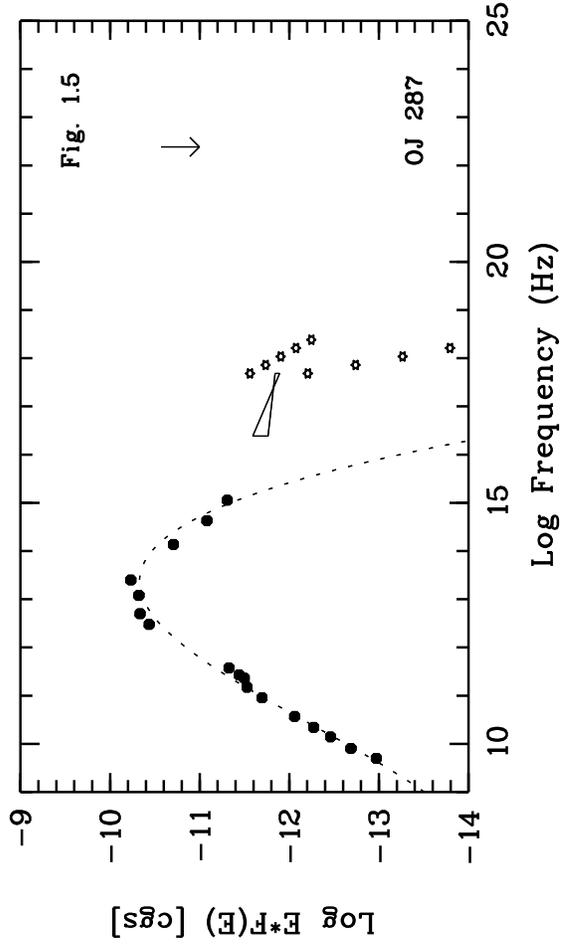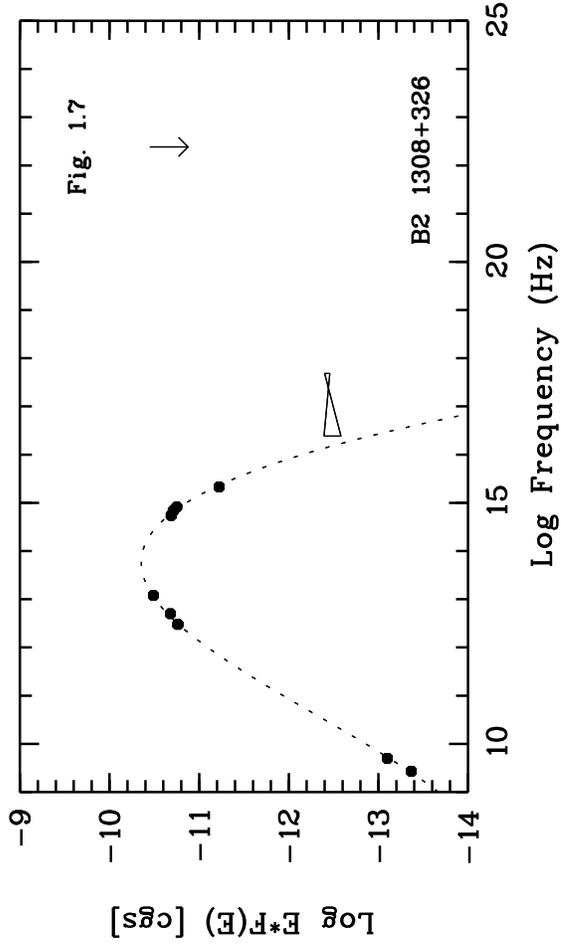

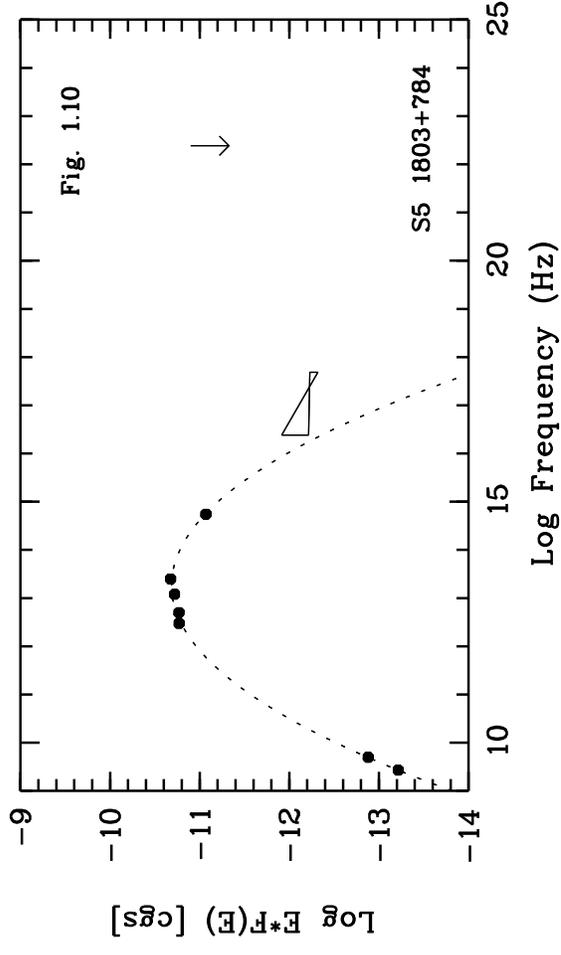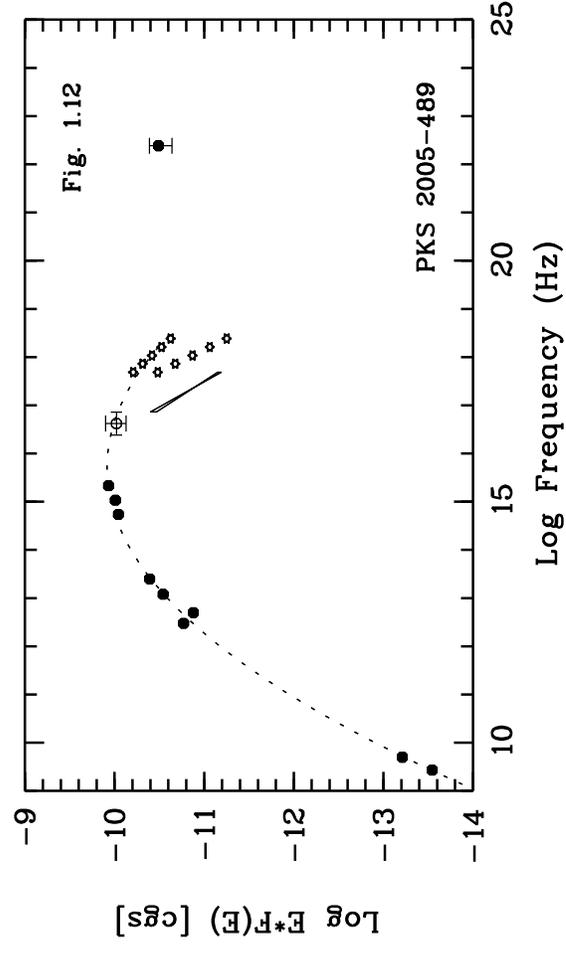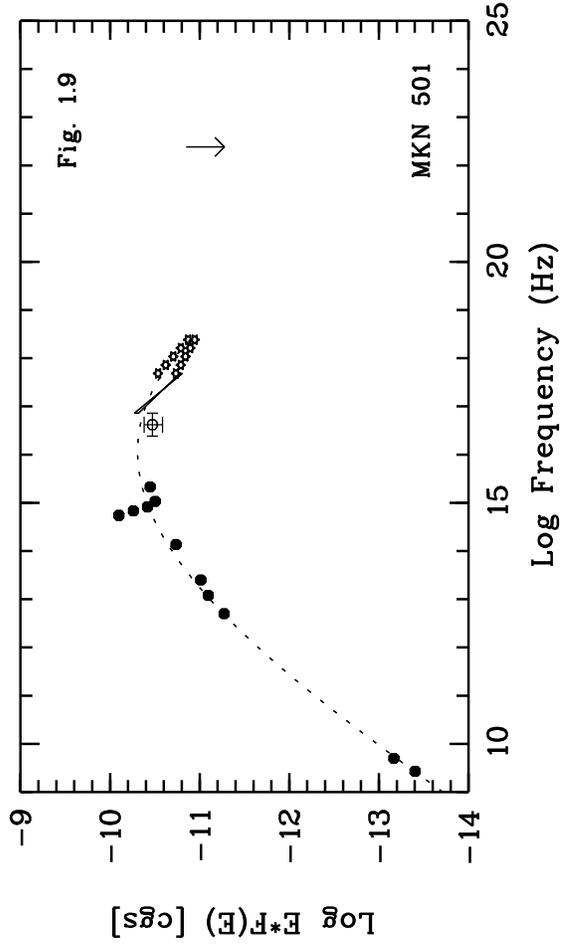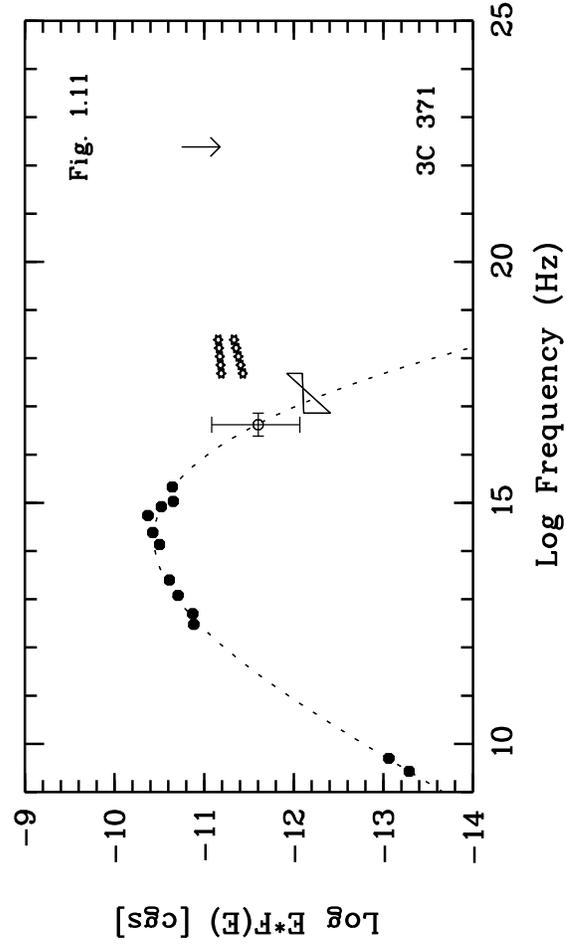

# ROSAT OBSERVATIONS OF RADIO-SELECTED BL LAC OBJECTS

A. Comastri[1,2], S. Molendi[2,3] & G. Ghisellini[4]

June 26, 1995

[1] Osservatorio Astronomico Bologna, via Zamboni 33, I–40126 Bologna, Italy
[2] Max–Planck–Institut für extraterrestrische Physik, D–85740 Garching, FRG
[3] Istituto di Fisica Cosmica del CNR, via Bassini 15, I–20133 Milano, Italy
[4] Osservatorio Astronomico Torino, Strada Osservatorio 20, I–10025 Pino Torinese, Italy

**Abstract**

ROSAT observations of a homogeneous sample of radio–selected BL Lac objects are presented. Results of a detailed spectral analysis in the soft $\sim 0.1 - 2.0$ keV energy range are discussed and compared with similar previously published results. The X–ray spectral shape is discussed in relation to the overall energy distribution with particular emphasys on the high energy $\gamma$–ray emission recently detected by CGRO–EGRET for about half of the objects in the sample. Dividing the objects in our sample on the basis of the radio to X–ray flux ratio ($\alpha_{rx}$) we show that objects with $\alpha_{rx} > 0.75$ have flatter X–ray spectra and are likely to be stronger $\gamma$–ray emitters than objects with $\alpha_{rx} < 0.75$. Moreover we note that the value of the peak energy of the synchrotron component, in a $\nu f_\nu$ plot, correlates with $\alpha_{rx}$ and $\alpha_x$.

## 1 INTRODUCTION

BL Lac objects are Active Galactic Nuclei characterized by compact radio cores with flat or inverted spectra, displaying superluminal motion on VLBI maps, relatively high optical and radio polarization, rapid variability in at least one of the radio, optical and X–ray bands. Their optical spectra are featureless, displaying weak (EW $\leq 5\text{Å}$) or no emission lines. For this reason they have been discovered (with very few exceptions) only at radio or X–ray frequencies. This has led to the subdivision of the BL Lac objects into two classes: RBL for radio–selected objects and XBL for the X–ray selected ones. It has been reported that RBL show more extreme properties than XBL, being more core–dominated, highly polarized, variable and luminous at radio and optical frequencies (Padovani 1992; Perlman & Stocke 1993; Jannuzi et al. 1994). Their multiwavelength spectra from radio to X–rays are well fitted by featureless non–thermal continua which are flat in the radio band and progressively steepen at higher energies (Giommi et al. 1995). The two classes show a different distribution in the $\alpha_{ro} - \alpha_{ox}$ diagram which is indicative of a different broad band energy distribution (Ledden & O'Dell 1985).

Some BL Lac objects are also strong $\gamma$–ray emitters with luminosities above 100 MeV of the order of $10^{46-48}$ ergs s$^{-1}$, if isotropic (Fichtel et al. 1994). In a few cases the emission is strongly variable on time scales as short as a day suggesting an extremely compact emission region.

The current interpretation of these violent characteristics and broad band energy distribution involves relativistic bulk motion of the emitting particles flowing in a jet. The emission mechanism from radio to ultraviolet frequencies is thought to be synchrotron radiation, while in the X–ray band the contribution of an inverse Compton component could be important (Synchrotron Self–Compton, hereafter SSC). A simple application of the SSC theory (Ghisellini et al. 1993) reveals that most of the BL Lac objects are relativistically beamed.

In the relativistic beaming scenario the BL Lac objects are thought to be low luminosity Fanaroff–Riley type I radio galaxies whose emission is dominated by the relativistic jet aligned with the line of sight (Urry & Padovani 1994). In this scenario the more extreme properties of the RBL class are explained as due to a smaller angle between the jet direction and the line of sight. The similar luminosities of the two classes in the X–ray band can be explained assuming that the X–ray beam is broader than the radio/optical one (Ghisellini & Maraschi 1989, Celotti et al. 1993). In this case the XBL objects should outnumber the RBL ones. This model accounts for the different break energies in the synchrotron spectrum of XBL and RBL.

Padovani & Giommi (1995), on the other hand, have proposed that the difference in break energies is an intrinsic property of XBL and RBL and is not due to a different viewing angle. In their scenario the BL Lac population is unique and the XBL type represents the tail of a broad distribution in the overall spectral shapes (i.e. the objects where the synchrotron component is dominant up to the X–rays domain).

The soft X–ray band lies where both the synchrotron and self–Compton component contribute to the total emitted power. Therefore a study of the soft X–ray properties is quite important to understand the physics of these extreme objects.

Worral & Wilkes (1990), from the analysis of Einstein IPC data have shown that the mean XBL and RBL spectra are not significantly different. A similar result has been found by Sambruna et al. (1994) using EXOSAT ME + LE. Finally Ciliegi, Bassani & Caroli (1995) using the archival data of various satellites (*Einstein* IPC and MPC, HEAO–1 A2, EXOSAT LE and ME) suggest that the X–ray spectra of the two classes are different, XBL showing convex X–ray spectra (the spectrum steepens toward high energies) and RBL concave spectra (the spectrum flattens at higher frequencies), although their statistics are somewhat poor.

We point out that in all the above studies RBL are underepresented as a consequence of their weaker X–ray flux. The very high sensitivity of ROSAT in the soft X–ray band allows us, for the first time, to perform a detailed analysis of an RBL sample.

The outline of the paper is as follows: in §2 the analysis of PSPC observations of a well defined sample of RBL type objects is described in some detail; in §3 the multiwavelength properties and correlations among the various band are discussed; summary and conclusions are presented in §4.

# 2 ROSAT observations

## 2.1 The sample

The sample has been selected from the catalogue of radio bright ($S_{5\ GHz} > 1$ Jy) BL Lacs objects of Stickel et al. (1991). In order to obtain high quality soft X–ray spectra we have considered only the optical bright ($m_V < 17$) objects, with the additional requirement of a relatively low line of sight column density ($N_H < 6 \times 10^{20}$ atoms cm$^{-2}$). Twelve BL Lacs satisfy this selection criterion. A summary of the properties of the objects in our sample is reported in Table 1.

## 2.2 Data reduction

The data reported here are from observations carried out with the ROSAT X–ray telescope (Trümper 1983) with the Position Sensitive Proportional Counter (PSPC–B) in the focal plane (Pfeffermann et al. 1986). The PSPC provides a bandpass in the range $\sim 0.1 - 2.4$ keV over a $\sim 2$ degrees diameter field of view. The energy resolution is $\Delta E/E = 0.43(E/0.93)^{-0.5}$ (FWHM) with E in keV. All the observations were performed in the standard "wobble" mode. In this mode the detector is moved back and forth from the pointing direction by $\sim 3'$ with a $\sim 400$ s period. This was done in order to smooth the detector structure and prevent the shadowing of X–ray sources behind the struts and the wire grid of the PSPC entrance window. The ROSAT observation log for our sample is shown in Table 2. The reported count rates are in pulse invariant (PI) channels and are background subtracted. The lowest channel used was PI channel 11, as the detector response matrix is not calibrated below this channel. The highest channel depends on the source spectrum and the signal to noise ratio of the observation, ranging from 200 to 240. Above channel 240 the XRT effective area is not well calibrated. For most of the sources, observed during the PV phase and/or the AO–1, the DRM06 (released on March 1992) detector response matrix was used, while for the other sources, observed during the AO–2, the last release of the detector response matrix was adopted (DRM36, released on January 1993).
The photon event files were analyzed using the EXSAS/MIDAS software (version 94JAN, Zimmerman et al. 1993), and the extracted spectra were analyzed using version 8.33 of XSPEC (Shafer et al. 1991).

## 2.3 Spectral analysis

### 2.3.1 Single Power law fits

The source spectra were rebinned in order to obtain a signal to noise ratio (S/N) $\geq 5$ for each bin thus enabling the application of the $\chi^2$ statistic. The data were then compared with various trial models. The results of fitting a single power law model with low energy absorption due to cold material with solar abundances (Morrison & McCammon 1983) are given in Table 3. The reported errors are 90% confidence intervals on two interesting parameters (Lampton, Margon & Bowyer 1976).

In a first run all the spectra were fitted with a power law model with the column density fixed at the Galactic value (Elvis, Lockman & Wilkes 1989 when available, Dickey & Lockman 1990 otherwise). While for most of the objects the spectra are adequately described by the model, for three sources the fits are unacceptable, in the sense that the probability associated to the reduced $\chi^2$ is smaller than 1%. The range of measured values of energy spectral indices is large, going from $\simeq 0.2$ to $\simeq 2.2$. The mean for the whole sample has been computed using a maximum likelihood algorithm (see Maccacaro et al. 1988) which has the advantage, over a simple algebraic mean, of weighting the individual energy indices according to their measured errors. The algorithm assumes that the intrinsic distribution of energy indices is Gaussian and allows the determination of the mean $\alpha_p$, the intrinsic dispersion $\sigma_p$ and the respective errors. The results are as follows: $\alpha_p = 1.30 \pm 0.25$, $\sigma_p = 0.49^{+0.23}_{-0.14}$ where the confidence intervals are at 90% level for one interesting parameter. The fact that $\sigma_p$ is not consistent with zero at a high significance level implies that the intrinsic distribution of spectral indices is not consistent with a single value.

In a second run all parameters were free to vary. In this case all the objects are well fitted (Table 3). The average properties as determined using the maximum likelihood algorithm are: $\alpha_p = 1.26 \pm 0.20$, $\sigma_p = 0.36^{+0.20}_{-0.11}$, similar to those derived for $N_H$ fixed at the Galactic value.

In order to investigate the origin of the difference between the fitted and the Galactic values of $N_H$ we have computed the $\Delta N_H$ distribution, defined as $\Delta N_H \equiv N_H^{x-ray} - N_H^{gal}$. The errors on $\Delta N_H$ for the individual objects have been computed by summing in quadrature the errors on $N_H^{x-ray}$ and $N_H^{gal}$, where $\sigma_{N_H^{gal}}$ is $10^{19}$ cm$^{-2}$ for the objects observed by Elvis, Lockman & Wilkes (1989) and $10^{20}$ cm$^{-2}$ for all the others. The mean and the intrinsic dispersion for $\Delta N_H$ have been derived using the maximum likelihood algorithm described above. The results are as follows: $(\Delta N_H)_p = (-0.45 \pm 0.48) \times 10^{20}$ cm$^{-2}$ and $(\sigma_{\Delta N_H})_p = 0.80^{+0.43}_{-0.28} \times 10^{20}$ cm$^{-2}$. The fact that the mean is consistent with zero indicates that on average a single power law spectrum, with $N_H$ fixed at the Galactic value, is a good description of the data. However the presence of an intrinsic dispersion significantly different from zero implies that for some sources a more complex model is required, this is also suggested by the unacceptably high values of reduced $\chi^2$ for the $N_H \equiv N_{H\ Gal.}$ fits of MKN 501, S5 0716+714 and PKS 2005−489. We note that objects characterized by $N_H^{x-ray} > N_H^{gal}$ can easily be explained in terms of intrinsic absorption, while objects with $N_H^{x-ray} < N_H^{gal}$ require a two component model for the X-ray spectrum. Consequently we have decided to fit two component models to all the objects for which $N_H^{x-ray}$ is significantly different from $N_H^{gal}$.

## 2.4 Two component models

We have fitted two component models to all sources for which the $N_H^{x-ray}$ is significantly different from $N_H^{gal}$ (see column 6 in Table 3). In all but one case (MKN 501) $N_H^{x-ray} < N_H^{gal}$, indicating a steepening of the spectrum toward low energies. A double power law model has been fitted to these objects and results are reported in Table 4. In order to better constrain the soft component parameters the high energy power law slope has been fixed at the values found from higher energies observations (Sambruna et al. 1994; Worral & Wilkes 1990). We point out that for most of the objects the improvement with respect to the single power law with $N_H \equiv N_H^{gal}$ is statistically significant. This is clearly indicated by the F-test values reported in Table 4. Note that for PKS 0537−441, the significance of the improvement is only marginal.

In five objects the power law dominating at soft energies is steeper than the one dominating at high energies; while for MKN 501 the opposite is true, i.e. the power law dominating at soft energies is flatter than the one dominating at high energies. On the basis of the analysis presented in this and in the previous subsection it appears that for some of the objects in our sample the soft X–ray spectrum can only be well described by a two component model (i.e. a double power law plus Galactic absorption). Given that the objects in Table 4 are the ones with highest statistics in our sample, it is tempting to speculate that deeper observations of the remaining objects will also reveal a more complex spectral shape. The physical implications of these results will be discussed in Section 3.

## 2.5 Temporal analysis

The background subtracted light curves have been computed for each source. There is no evidences of large amplitude flux variability, except for S5 0716+714 which is discussed in a separate paper (Cappi et al. 1994). In that paper the observed flux and spectral variability is explained in the framework of the SSC model.

# 3 Results

## 3.1 X–ray spectral results: comparison with other samples

The spectral analysis results have been compared with the available X–ray data in the literature. ROSAT data have been collected from the All Sky Survey and from some pointed observations (Brinkmann & Siebert 1994; Brunner et al. 1994; Maraschi et al. 1995; hereafter BBM). We have not considered the few objects in common with our sample however the individual spectra reported by these authors are in good agreement with ours. The X–ray spectral data from other missions have been retrieved from Worral & Wilkes (1990) for the *Einstein* IPC, from Sambruna et al. (1994) for the EXOSAT ME and from the compilation of Ciliegi et al. (1993).

A maximum likelihood analysis (cfr. section 2.3), gives the results reported in Table 5, for the best fit mean slope and 90% confidence errors. A good agreement between the mean spectral indices in the soft band is found (see column 4), despite the fact that while the BBM sample is dominated by XBL objects, the *Einstein* sample contains mostly RBL objects and our sample contains only RBL. A close similarity between XBL and RBL X–ray spectra was already pointed out by Worral & Wilkes (1990), from the analysis of the *Einstein* IPC data. The same average spectral index, within the errors, is also found in the 2 to 10 keV band.

We would like to point out that the division between XBL and RBL in our and other samples refers to the energy band in which the objects were originally discovered. However a physically more meaningful division can be obtained by considering the radio to X–ray spectral index, defined as $\alpha_{rx} = log(F_r/F_x)/(\nu_x/\nu_r)$, where $\nu_r$ = 5 GHz and $\nu_x$ = 1 keV. (cfr. Ghisellini et al 1986, Padovani and Giommi 1995). Following Padovani and Giommi (1995) we define as Steep Radio to X–ray objects (SRX) those for which $\alpha_{rx} > 0.75$ and Flat Radio to X–ray objects (FRX) those for which $\alpha_{rx} < 0.75$, without applying any K–correction for the monochromatic fluxes.

The results (column 6 and 8 of Table 5) indicate that the spectral indices of SRX objects are significantly flatter than the FRX ones. Note that this is true also at medium (2 − 10 keV) energies (last row of the table), even if the statistic for SRX objects is rather poor.

These results can be understood in the framework of the inhomogeneous jet model for the X–ray emission of BL Lac objects (Ghisellini et al. 1985). In this model SRX X–ray spectra are expected to be relatively flat, being already dominated by the inverse Compton component, while FRX X–ray spectra should be steeper, because their X–ray emission is mainly due to the synchrotron component. Thus it seems that a division of BL Lac objects based on the Radio to X–ray flux ratio is more meaningful than one based on the energy band in which the object was originally discovered.

In order to further investigate this effect the X–ray data of our sample have been compared with the available multiwavelength data from radio to $\gamma$–ray energies.

## 3.2 Broad Band Energy Distributions

For each object multiwavelength data have been collected from the literature. Radio fluxes at 5 GHz are from the 1 Jy all–sky catalogue of Kühr et al. (1981), while radio fluxes at 2.7 GHz together with optical magnitudes and redshifts have been collected from the Veron & Veron (1993) catalogue. Ultraviolet fluxes at 1400 and 2800 Å have been taken from Pian & Treves (1993). Far–infrared fluxes have been retrieved from the Impey & Neugebauer (1988) compilation of IRAS data. Soft X–ray data are from the present work. The hard (2 − 10 keV) X–ray data, when available, have been collected from the compilation of the EXOSAT results by Sambruna et al. (1994). The high energy $\gamma$–ray data are from the CGRO-EGRET all–sky survey source catalogue (Fichtel et al. 1994 and Von Montigny et al. 1995).

The broad band energy distributions (hereafter BBED) for each source in the $\nu - \nu F_\nu$ representation are shown in Fig. 1.1 to 1.12. When more than one measurement for a given energy is available we have considered the maximum flux. This should minimise variability effects, known to be important for these sources.

The BBED of all the objects is characterized by a smooth rise between the radio and IR band followed by a cutoff at higher energies. In order to obtain a reliable estimate of the cutoff energy the BBED of each object has been fitted with a polynomial function of the type : $log\ [\nu F(\nu)] = a + b\ log\ \nu + c\ [log\ \nu]^2 + d\ [log\ \nu]^3$.

The peak energies are distributed over a wide range from the far infrared to the UV–soft–X–rays, ($\nu_{peak} \sim 10^{13} - 10^{16}$ Hz). A significant correlation has been found between $\nu_{peak}$ and $\alpha_{rx}$, Figure 2, and between $\nu_{peak}$ and $\alpha_x$, Figure 3, (both at > 99 % using a non–parametric Spearman rank test).

The data used to construct the BBED are not simultaneous and so should be considered with caution. Nevertheless the wide range of cutoff energies coupled with the correlation between $\alpha_{rx}$ and the peak energy of the synchrotron component confirms the Padovani & Giommi (1995) results, suggesting a physical subdivision of the BL Lac objects on the basis of $\alpha_{rx}$.

## 3.3 The $\gamma$–ray bright objects

Five objects of our sample have been detected at $\gamma$–ray energies (> 100 MeV) by the EGRET detector on board CGRO. It is interesting to note that the X–ray spectra of four of the five objects (cfr. § 2.4) can be fitted with a double power law component which flattens towards high energies while the remaining object (S4 0954+658) has the flattest X–ray spectrum of the sample ($\alpha_x = 0.23$). There is one more object (3C 371) that can be fitted with a double power law component, but it has not yet been detected by EGRET. If the presence of a hard tail in the ROSAT X–ray spectrum suggests the presence of the self–Compton component, then 3C 371 is a good candidate to be detected by EGRET, when in a 'high state'. In fact the large amplitude variability could easily make this source undetectable by EGRET during quiescent phases or 'low states'.

The origin of the $\gamma$–ray emission is, at present, still unclear, even if there is increasing consensus towards the idea of inverse Compton emission being responsible of the high energy radiation. The seed photons to be comptonized could originate from synchrotron emission localized in the same region producing the $\gamma$–rays (i.e. SSC models; Maraschi, Ghisellini & Celotti 1992), or could be produced externally, by the accretion disk (Dermer, Schlickeiser & Mastichiadis 1992), by the broad line region or by some scattering material surrounding a relativistically moving blob (Sikora, Begelman & Rees, 1994, Blandford 1993, Blandford & Levinson 1995).

In the case of BL Lac objects, the contribution of the external photons is probably less important than in other $\gamma$–ray bright objects, due to the lack of broad emission lines and of any observable thermal emission features. This leaves the SSC mechanism as the most likely process for the origin of the high energy emission (see also Ghisellini & Maraschi 1994).

In this framework the flux level and the spectral shape in the X–ray band is related to the $\gamma$–ray emission: flat X–ray spectral indices indicate an important self Compton contribution and therefore a bright $\gamma$–ray source; steep X–ray spectra, smoothly connecting with the optical–UV, are probably synchrotron emission: in this case the Compton component is relatively less important, and the source should be less bright also in the $\gamma$–ray band.

As a consequence the brightest $\gamma$–ray BL Lacs should be the SRX ones, because of their stronger Compton emission.

This is in apparently not in complete agreement with our findings, because 2 out of 5 $\gamma$–ray bright BL Lacs in our sample (S5 0716+714 and PKS 2005–489), have an overall spectrum of the FRX type. However we point out that in the large sample of FRX sources (20 objects) of Brinkmann et al. (1994), there is no source detected by EGRET.

Regarding S5 0716+714, we also note that it is an intermediate object whose $\alpha_{rx}$ is steeper or flatter than 0.75 according to the X–ray flux level (cfr. Fig. 2).

Concerning PKS 2005–489 we point out that it is one of the closest objects in our sample ($z = 0.071$). This is the reason why it is detected by EGRET even though its $\gamma$–ray to optical flux ratio is the smallest among the sources in our sample.

For the remaining 7 objects in our sample an upper limit is available in the EGRET band. For all these objects the extrapolation of the X–ray power law falls below the EGRET upper limit. The object for which the extrapolation of the power law comes closer to the upper limit is 3C 371. This object is therefore a good candidate for further EGRET observations.

### 3.3.1 The case of S4 0954+658

The BL Lac object S4 0954+658 deserves a special attention because of the following peculiar characteristics:

i) it has the flattest X–ray spectrum among the sources of our sample;
ii) The peak of its synchrotron spectrum is at very low energies ($\nu_p \sim 10^{13}$ Hz), while

the $\gamma$–ray spectral index indicates that most of the high energy luminosity is emitted above 1 GeV.

iii) The ratio of the $\gamma$–ray to the synchrotron luminosity exceeds $\sim 10$, and is the largest for the sources in our sample. Such large values are typical for highly polarized quasar and flat spectrum radio quasars, but extreme for BL Lacertae objects (Dondi & Ghisellini, 1995)

Even if points i) and ii) above can be weakened by variability, it is reasonable to question for this source the applicability of a simple SSC model.

In the framework of this model, the same electrons responsible for the peak in the synchrotron emission (at the frequency $\nu_s$) are the ones dominating the contribution at the frequency $\nu_c$ where the Compton emission peaks. If the Lorentz factor of these electrons is $\gamma_b$, we have $\gamma_b = (\nu_c/\nu_s)^{1/2} \sim 3 \times 10^5 (\nu_{c,24}/\nu_{s,13})^{1/2}$, where $\nu_s = 10^{13}\nu_{s,13}$ and $\nu_c = 10^{24}\nu_{c,24}$. Since $\nu_s = 2.8 \times 10^6 \, B \, \delta \, \gamma_b^2$ Hz, where $B$ is the magnetic field in Gauss and $\delta$ is the beaming factor, we derive $B \, \delta = 3.6 \times 10^{-5}$. For values of $\delta$ near 10 or more, the derived magnetic field is so small that even the cosmic background radiation becomes more important than synchrotron radiation in providing the seed photons for Compton scattering.

We therefore suggest that in this source the main mechanism at the origin of the $\gamma$–ray emission is not SSC.

An attractive alternative is the model of Sikora, Begelman & Rees (1994), which can easily explain: the large energy separation between the peaks of the synchrotron and $\gamma$–ray emission, the dominance of the high energy power and the very flat X–ray spectrum.

Obviously the difficulty of this model is the lack, in S4 0954+658, of any thermal emission, indicating the possible relevance of the seed photons produced externally to the $\gamma$–ray emitting region. In fact, in the observation of Stickel et al. (1993) the optical spectrum of this source reveals the presence of only a very weak forbidden [O II] line. However, it should be interesting to observe again this source in low optical states, to find the possible presence of broad emission lines, albeit weak. In this case S4 0954+658 could be classified as an intermediate BL Lac – HPQ type, with weak emission lines, (which however give an important contribution to the seed photons to be Comptonized), generally swamped by the beamed non–thermal emission.

## 4 Summary and Conclusions

We have analyzed ROSAT PSPC data for 12 BL Lac objects drawn from the complete sample of RBL of Stickel et al. (1991) with the further requirements of a bright optical magnitude and a relatively low column density on the line of sight.

The mean X–ray spectrum, adopting a single power law model, is relatively steep ($\alpha_x \simeq 1.3$), the X–ray spectral indices distribution is characterized by a large dispersion with energy indices in the range 0.2 to 2.2. Our average $\alpha_x$ is consistent with values derived from previous studies (cfr. Table 5 column 4). Dividing the objects of all samples according to their radio to X–ray flux ratio and defining FRX those with $\alpha_{rx} < 0.75$ and SRX those with $\alpha_{rx} > 0.75$, we find that SRX have a significantly flatter X–ray spectral index than FRX.

For six objects of our sample a double power law provides a better description of the data. In all, but one case, the spectrum flattens towards high energies. For MKN 501, an FRX object, the spectrum steepens with increasing energy.

The BBED of each object can be described by a relatively smooth power law spectrum which progressively steepens from the radio to the X-ray band. This suggests a single emission mechanism, i.e. synchrotron radiation over a broad energy range. A careful analysis of the BBED indicates a large spread in the frequency $\nu_{peak}$ where the synchrotron component peaks in a $\nu - \nu F(\nu)$ plot.

We have shown that $\nu_{peak}$ is correlated with $\alpha_{rx}$ and $\alpha_x$ for the objects in our sample suggesting that:

1) in objects with large $\nu_{peak}$ the synchrotron component extends up to the ROSAT band, giving rise to a steep $\alpha_x$. On the contrary when the synchrotron spectrum peaks at low energies the X-ray emission is dominated by the self Compton process yielding a flat spectral index;

2) the X-ray spectra of SRX objects are dominated by the flat self Compton component while the steep X-ray spectra of FRX are still dominated by the synchrotron emission.

As a consequence both $\nu_{peak}$ and $\alpha_{rx}$ can be used to divide BL Lac objects into the two classes FRX and SRX. The above correlations can be adequately explained both by the models of Padovani & Giommi (1995) and of Ghisellini and Maraschi (1989), but cannot be used to discriminate between the two.

Five objects of our sample have been detected by EGRET above 100 MeV. We have shown that 3 of these are SRX objects whose flat X-ray spectrum indicates the importance of the self Compton mechanism. Of the remaining 2, one (S5 0716+714) is of intermediate type and the other (PKS 2005-489) is probably detected only because of its proximity.

For all the remaining objects not detected by EGRET the extrapolation of the X-ray power law falls below the 100 MeV upper limit. 3C 371 is the object for which the extrapolation comes closer to the EGRET upper limit and is consequently a good candidate for further observations by CGRO.

Broad band simultaneous X-ray observations as provided by ASCA (0.4 − 10 keV) and SAX (0.5−100 keV) are clearly required to further constrain the relative importance of the different emission mechanisms and the models for the origin of $\gamma$-ray emission.

# Figure Captions

**FIG. 1.1–1.12**   The Radio to $\gamma$–ray Broad Band Energy Distributions for the whole sample in the $\nu F(\nu)$ form (units of ergs cm$^{-2}$ s$^{-1}$). Filled circles represent radio to ultraviolet fluxes collected from the literature (see text for details), stars represent the best fit $(2-10$ keV) X–ray slopes from EXOSAT ME (Sambruna et al. 1994). When more than one measure was available the highest and the lowest intensity spectra are reported. ROSAT results are represented as solid bow–tie shapes in the $0.3-2.0$ keV energy range for the double power law objects (cfr. Table 4) and in the $0.1-2.0$ keV range for the others. The empty circles in the $0.1-0.3$ band represent the soft component fit results for the double power law objects. The CGRO–EGRET results are downward pointing arrows at 100 MeV if an upper limit is available, filled points with relative $1\sigma$ errors at 100 MeV if a detection is available, and solid bow–tie contours in the $0.1-1.0$ GeV range if a $\gamma$–ray spectrum is available. The dotted line over the frequency range $log\ \nu = 9-19$ is the polynomial fit to the radio to UV or X–ray data. We note that this fit is not intended to reproduce in detail the Broad Band Energy Distribution, but rather to estimate the value of the peak of the distribution. Note that for PKS 0735+178 (Fig. 1–4) the stars refer to the IPC spectrum of Worrall & Wilkes (1990), while the $\gamma$–ray detection above 100 MeV is only at $3\sigma$ (Mattox 1994, private communication). For S5 0716+714 (Fig. 1–3) we have reported the time resolved X–ray spectral analysis results (see Cappi et al. 1994 for details). The low intensity state is reported with solid contours, while the high intensity one with dashed contours. The same notation have been used for the two $\gamma$–ray spectra.

**FIG. 2**   The $\alpha_{rx}$ vs. $\nu_{peak}$ diagram for the objects of our sample. The three FRX type ($\alpha_{rx} < 0.75$) objects are labeled. Note the intermediate character of S5 0716+714, which can be considered an FRX object when in the high state or an SRX one when in the low state.

**FIG. 3**   The $\alpha_x$ vs. $\nu_{peak}$ diagram for the objects of our sample.

## Table 1: The sample

| Name | $RA(1950)$ | $DEC(1950)$ | $F_R^a$ | $m_V^b$ | $z$ | $N_{HGal}^c$ |
|---|---|---|---|---|---|---|
| S5 0454+844 | 04 54 57.4 | 84 27 53 | 1.398 | 16.5 | 0.112 | 5.74 |
| PKS 0537−441 | 05 37 21.0 | −44 06 45 | 3.960 | 16.5 | 0.896 | 4.77 |
| S5 0716+714 | 07 16 12.9 | 71 26 15 | 1.121 | 15.5 | >0.3 | 3.95 |
| PKS 0735+178 | 07 35 14.1 | 17 49 09 | 1.990 | 16.2 | >0.424 | 4.35 |
| OJ 287 | 08 51 57.2 | 20 17 58 | 2.610 | 15.4 | 0.306 | 2.75 |
| S4 0954+658 | 09 54 57.8 | 65 48 15 | 1.460 | 16.7 | 0.368 | 3.71 |
| B2 1308+326 | 13 08 07.6 | 32 36 40 | 1.590 | 15.2 | 0.996 | 1.10 |
| OQ 530 | 14 18 06.2 | 54 36 57 | 1.090 | 15.7 | 0.152 | 1.18 |
| MKN 501 | 16 52 11.7 | 39 50 25 | 1.313 | 13.9 | 0.034 | 1.73 |
| S5 1803+784 | 18 03 39.2 | 78 27 55 | 2.633 | 16.4 | 0.684 | 5.00 |
| 3C 371 | 18 07 18.5 | 69 48 58 | 1.740 | 14.2 | 0.051 | 4.67 |
| PKS 2005−489 | 20 05 46.6 | −48 58 43 | 1.190 | 13.4 | 0.071 | 4.96 |

NOTES:
[a] Radio flux at 5 GHz in Jy (Kühr et al. 1981).
[b] Visual optical magnitude (Veron & Veron 1993)
[c] Galactic column density (units of $10^{20}$ cm$^{-2}$). For OJ 287, PKS 0735+178, MKN 501 and S5 1803+784 the values are from Elvis et al. (1989). For all the other objects from Dickey & Lockman (1990).

## Table 2: ROSAT PSPC Observations of BL Lacs

| Name | Date | exposure (s) | count rate (cts s$^{-1}$) |
|---|---|---|---|
| S5 0454+844 | 04-APR-1991 | 3358 | $0.016 \pm 0.003$ |
| PKS 0537−441 | 10-APR-1991 | 2598 | $0.355 \pm 0.013$ |
| S5 0716+714 | 08-MAR-1991 | 21043 | $0.802 \pm 0.006$ |
| PKS 0735+178 | 27-OCT-1992 | 6684 | $0.089 \pm 0.004$ |
| OJ 287 | 16-APR-1991 | 3622 | $0.273 \pm 0.009$ |
| S4 0954+658 | 16-APR-1991 | 6772 | $0.052 \pm 0.003$ |
| B2 1308+326 | 23-JUN-1991 | 8386 | $0.095 \pm 0.004$ |
| OQ 530 | 19-JUL-1990 | 11516 | $0.222 \pm 0.005$ |
| MKN 501 | 24-FEB-1991 | 7649 | $6.950 \pm 0.031$ |
| S5 1803+784 | 07-APR-1992 | 6911 | $0.081 \pm 0.004$ |
| 3C 371 | 09-APR-1992 | 10461 | $0.122 \pm 0.004$ |
| PKS 2005−489 | 27-APR-1992 | 11487 | $2.780 \pm 0.016$ |

## Table 3: Single Power Law Spectral Fits

| Name | $N^a_{H gal.}$ | $\alpha^b$ | $F^c$ | $\chi^{2(d)}_\nu$ | Free $N^e_H$ | $\alpha^b$ | $\chi^{2(d)}_\nu$ |
|---|---|---|---|---|---|---|---|
| S5 0454+844 | 5.74 | $1.47^{+0.71}_{-0.97}$ | 0.05 | ... | ... | ... | ... |
| PKS 0537−441 | 4.11 | $1.52^{+0.10}_{-0.10}$ | 0.82 | 1.75/19 | 2.63 (1.70 − 3.70) | $1.07^{+0.35}_{-0.34}$ | 1.37/18 |
| S5 0716+714 | 3.95 | $2.09^{+0.04}_{-0.05}$ | 1.26 | 2.35/73 | 2.93 (2.71 − 3.15) | $1.73^{+0.08}_{-0.09}$ | 1.20/72 |
| PKS 0735+178 | 4.35 | $1.25^{+0.16}_{-0.17}$ | 0.24 | 1.09/8 | 3.04 (1.64 − 4.28) | $0.89^{+0.50}_{-0.47}$ | 0.82/7 |
| OJ 287 | 2.75 | $1.14^{+0.09}_{-0.08}$ | 0.62 | 0.75/11 | 3.06 (2.11 − 4.14) | $1.25^{+0.37}_{-0.35}$ | 0.78/10 |
| S4 0954+658 | 3.71 | $0.23^{+0.24}_{-0.26}$ | 0.17 | 0.79/7 | 7.19 (2.42 − 32.22) | $0.67^{+1.55}_{-0.74}$ | 0.6/6 |
| B2 1308+326 | 1.10 | $0.95^{+0.10}_{-0.09}$ | 0.15 | 1.14/10 | 1.50 (0.74 − 2.46) | $1.12^{+0.41}_{-0.36}$ | 1.14/9 |
| OQ 530 | 1.18 | $1.04^{+0.05}_{-0.05}$ | 0.32 | 1.78/10 | 1.21 (0.81 − 1.69) | $1.06^{+0.21}_{-0.20}$ | 1.97/9 |
| MKN 501 | 1.73 | $1.35^{+0.03}_{-0.02}$ | 9.92 | 2.79/108 | 2.37 (2.26 − 2.49) | $1.60^{+0.05}_{-0.04}$ | 1.15/107 |
| S5 1803+784 | 5.00 | $1.17^{+0.14}_{-0.16}$ | 0.25 | 1.35/11 | 5.25 (3.53 − 7.36) | $1.22^{+0.45}_{-0.46}$ | 1.47/10 |
| 3C 371 | 4.67 | $1.22^{+0.10}_{-0.12}$ | 0.35 | 1.46/19 | 2.69 (1.82 − 3.65) | $0.69^{+0.29}_{-0.28}$ | 0.58/18 |
| PKS 2005−489 | 4.96 | $2.21^{+0.04}_{-0.03}$ | 5.21 | 2.53/91 | 4.00 (3.81 − 4.18) | $1.89^{+0.07}_{-0.06}$ | 1.23/90 |

NOTES:
[a] Galactic column density (units of $10^{20}$ cm$^{-2}$).
[b] Energy spectral index.
[c] Flux at 1 keV (in $\mu$Jy).
[d] Reduced $\chi^2$ and degrees of freedom.
[e] Column density obtained from the X-ray spectral fits (units of $10^{20}$ cm$^{-2}$).
For S5 0454+844 the number of detected counts was too low for an accurate spectral analysis.
The reported slope has been obtained using the hardness ratio technique with $N_H \equiv N_{HGal}$.
Errors are quoted at 90% confidence intervals for one interesting parameter ($\chi^2_{min} + 2.71$, column 3)
and for two parameters ($\chi^2_{min} + 4.61$, columns 6 and 7).

Table 4: Double Power Law Spectral Fits with $N_H = N_{HGal}$.

| Name | $N_{HGal}^a$ | $\alpha_1^b$ | $E_{break}^c$ | $\alpha_2^d$ | $F^e$ | $\chi_\nu^{2(f)}$ | $p^g(F-test)$ |
|---|---|---|---|---|---|---|---|
| PKS 0537−441 | 4.11 | $2.03^{+0.46}_{-0.37}$ | 1.01 | 0.30 | 0.39 | 1.54/18 | > 90% |
| S5 0716+714 | 3.95 | $2.67^{+0.12}_{-0.12}$ | 0.82 | 1.00 | 0.78 | 1.09/72 | > 99.9% |
| PKS 0735+178 | 4.35 | $2.57^{+1.82}_{-1.12}$ | 0.35 | 0.70 | 0.21 | 0.67/7 | > 97.5% |
| 3C 371 | 4.67 | $5.46^{+0.55}_{-2.22}$ | 0.22 | 0.95 | 0.34 | 0.82/18 | > 99.5% |
| PKS 2005−489 | 4.96 | $3.40^{+0.27}_{-0.23}$ | 0.30 | 1.70 | 4.65 | 1.22/90 | > 99.9% |
| MKN 501 | 1.73 | $0.43^{+0.22}_{-0.25}$ | 0.31 | 1.55 | 10.2 | 1.20/107 | > 99.9% |

NOTES:
[a] Galactic column density (units of $10^{20}$ cm$^{-2}$).
[b] Energy index of the soft component.
[c] Break energy in keV.
[d] Energy index of the hard component fixed at the best fit value of the EXOSAT ME (2–10 keV) observations (Sambruna et al. 1994):
For PKS 0735+178 and S5 0716+714, 2–10 keV spectra are not available.
The adopted values are from *Einstein* IPC (Worral & Wilkes 1990) and ROSAT PSPC (Cappi et al. 1994) respectively.
[e] Flux at 1 keV in $\mu$Jy
[f] Reduced $\chi^2$ and degrees of freedom.
[g] Significance of the improvement of a double component fit with respect to the single power law computed with the F-test.

Table 5: BL LACs Samples: mean X-ray spectral properties

| Sample | Energy Range | # obj. | $<\alpha_X>$ | # SRX | $<\alpha_X>$ | # FRX | $<\alpha_X>$ |
|---|---|---|---|---|---|---|---|
| This work | 0.1–2.4 | 12 | $1.26 \pm 0.20$ | 9 | $1.04 \pm 0.12$ | 3 | $1.74 \pm 0.12$ |
| $BBM^a$ | 0.1–2.4 | 21 | $1.48 \pm 0.14$ | 3 | $1.38 \pm 0.44$ | 18 | $1.49 \pm 0.14$ |
| $IPC^b$ | 0.2–4.0 | 25 | $1.10 \pm 0.13$ | 15 | $0.71 \pm 0.23$ | 10 | $1.26 \pm 0.12$ |
| Total soft | 0.1–4.0 | 58 | $1.30 \pm 0.10$ | 27 | $1.02 \pm 0.12$ | 31 | $1.48 \pm 0.10$ |
| $ME + GINGA^c$ | 2.0–10 | 24 | $1.34 \pm 0.13$ | 5 | $1.06 \pm 0.48$ | 19 | $1.40 \pm 0.11$ |

NOTES:

[a] Data collected from: Brinkmann & Siebert (1994); Brunner et al. (1994); Maraschi et al. (1995).
[b] Data collected from: Worral & Wilkes (1990); Ciliegi et al. (1993).
[c] Data collected from: Sambruna et al. (1994); Makino (1989)